\documentclass[twocolumn, switch]{article}
\usepackage{hyperref}	
\hypersetup{colorlinks = true,
            linkcolor = purple,
            urlcolor  = blue,
            citecolor = cyan,
            anchorcolor = black}
\usepackage[backend=biber,doi=false,isbn=false,url=false,eprint=false,style=nature]{biblatex}
\usepackage{preprint}
\usepackage[utf8]{inputenc}	
\usepackage[T1]{fontenc}	
\usepackage{booktabs} 		
\usepackage{nicefrac}		
\usepackage{microtype}		
\usepackage{lineno}		
\usepackage{lipsum}		
\usepackage{multicol}
\usepackage{graphicx}
\usepackage{float}			
\usepackage{stfloats}
\usepackage[labelsep=period]{caption}
\usepackage{lineno}
\usepackage{pdfpages}
\usepackage{color}
\usepackage{newfloat}
\usepackage{lipsum}
\usepackage{rotating}
\usepackage{titlesec}
\titlespacing\section{0pt}{12pt plus 3pt minus 3pt}{1pt plus 1pt minus 1pt}
\titlespacing\subsection{0pt}{10pt plus 3pt minus 3pt}{1pt plus 1pt minus 1pt}
\titlespacing\subsubsection{0pt}{8pt plus 3pt minus 3pt}{1pt plus 1pt minus 1pt}
\usepackage{subcaption}
\titleformat*{\section}{\large\bfseries}
\titleformat*{\subsection}{\normalsize\bfseries}

\title{Neural-Network Quantum States: A Systematic Review}

\usepackage{xwatermark}

\usepackage{authblk}

\author[1, 2]{David R. Vivas}
\author[1,2]
{Javier Madroñero}
\author[3]
{Víctor Bucheli}
\author[4]
{Luis O. González}
\author[1,2]
{John H. Reina}
\affil[ ]{
\texttt{david.vivas@correounivalle.edu.co} 
\vspace{15pt}}
\affil[1]{Centre for Bioinformatics and Photonics (CIBioFi), Universidad del Valle, Edificio E20 No.~1069, 760032 Cali, Colombia}
\affil[2]{Departamento de F\'isica, Universidad del Valle, 760032 Cali, Colombia}
\affil[3]{Escuela de Ingeniería de Sistemas y Computación, Universidad del Valle, 760032 Cali, Colombia}
\affil[4]{Departamento de Ingenier\'ia, Universidad Nacional de Colombia, Sede Palmira, Colombia}

\begin{document}


\twocolumn[ 
 \begin{@twocolumnfalse} 
\maketitle
\begin{abstract}
{\textbf{The so-called contemporary AI revolution has reached every corner of the 
social, human and natural sciences---physics included. In the context of quantum many-body physics, its intersection with machine learning has configured a high-impact interdisciplinary field of study; with the arise of recent seminal contributions that have derived in a large number of publications. One particular research line of such field of study is the so-called Neural-Network Quantum States, a powerful variational computational methodology for the solution of quantum many-body systems that has proven to compete with well-established, traditional formalisms. Here, a systematic review of literature regarding Neural-Network Quantum States is presented.}} 
\end{abstract}
\vspace{10pt}
\end{@twocolumnfalse}
]


\section{Introduction}

The exponential growth on the dimensionality of a quantum system in regard to its number of bodies impedes its classical, direct computational solution for anything more than small systems: in order to solve a 300-quantum bit (qubit) system with non-trivial interactions via diagonalization of its Hamiltonian it is required to solve a system of $2^{300}$ linear equations, a number larger than the estimated number of atoms in the universe\supercite{whittaker_eddingtons_1945}. Due to this curse of dimensionality, through the history of computational physics several computational methodologies such as the Variational ansatz Montecarlo\supercite{mcmillan_ground_1965} (VMC), the Density-Matrix Renormalization Group (DMRG) \supercite{white_density_1992, cirac_renormalization_2009, jahromi_universal_2019} or the Path-Integral Montecarlo formalism\supercite{barker_quantum-statistical_1979} have been developed to approach this problem. Nevertheless, the aforementioned methods are not exempt of limitations such as the lack of accurate ansatzs, the sign-problem or the specialization 
to one-dimensional configurations.

 In 2017, Carleo \& Troyer\supercite{carleo_solving_2017} developed a novel methodology based on the use of artificial neural networks (NNs) for the simulation of quantum many-body systems. This novel approach is based on the Variational Montecarlo technique, but uses as ansatz an adjustable parameter blackbox represented by a neural-network (Restricted Boltzmann Machine, RBM), thus eliminating the need of fine-tuning a hypothesis function modeled in terms of the properties of the ground-state, which are generally unknown in the lack of an analytical approximation or solution.

This approach, regarded as Neural-Network Quantum States (NQS) in the literature, has achieved state of the art results both in approximating the ground-state of large chains and lattices of quantum spins \supercite{sharir_deep_2020,hibat-allah_recurrent_2020} with an amount of variational parameters significantly lower than the used by methods such as DMRG \supercite{carleo_solving_2017, hibat-allah_recurrent_2020} and with the capability of executing unitary time evolution \supercite{carleo_solving_2017, carleo_unitary_2017, schmitt_quantum_2019, lopez-gutierrez_real_2019} and of reconstructing physical wavefunctions parting from experimental data \supercite{torlai_neural-network_2018}.

Several recent reviews\supercite{dunjko_machine_2018, jia_quantum_2019, carleo_machine_2019, melko_restricted_2019, carrasquilla_machine_2020, yang_neural_2020,Carrasquilla_Torlai_2021} encompass neural-network quantum states as a discussion topic: In \supercite{dunjko_machine_2018} several machine learning algorithms and their quantum physics applications are extensively discussed. In \supercite{jia_quantum_2019} a review of NQS representational power, relation to tensor networks and applications to condensed matter and quantum computing is presented. In \supercite{carleo_machine_2019} the authors describe the grand landscape of Machine Learning applications to several areas of the physical sciences. In \supercite{melko_restricted_2019}, the authors review and discuss the impact of Restricted-Boltzmann machines in quantum physics, with special emphasis on NQS and condensed matter. In \supercite{carrasquilla_machine_2020}, the author analyzes the grand landscape of the intersection between machine learning and quantum matter, with a section devoted to discussion on NQS. In \supercite{yang_neural_2020} the authors review NQS from a technical-focused point of view.  Lastly, in \supercite{Carrasquilla_Torlai_2021} the authors provide a practical guide for supervised phase transition learning and unsupervised quantum tomography and ground-state approximation by using different neural-network architectures. Regarding such works, this review focuses on providing the reader with a broad landscape of both the computational problem and the vast existing literature related to NQS since its development in 2017\supercite{carleo_solving_2017}.

Although a relatively novel methodology, the rapid growth of published works related to NQS research raises several questions: Which are the advancements and what is the current state of this methodology? What are the main challenges and perspectives in this line of research? In this work, we address such questions by means of a systematic review of literature that provides with a state-of-the art on Neural-Network Quantum States.

This article is structured as follows:
\begin{itemize}
\item{Section 1 contextualizes the reader on the computational problem, the method of interest and the open research questions up to date.}
\item{Section 2 describes the methodology executed for the literature search and presents the network of researchers obtained from this collection.}
\item{Section 3 describes and discusses the identified literature.}
\item{Section 4 presents the conclusions and perspectives reached during this review.}
\end{itemize}



\section{Methodology and systematic search of literature}

\begin{figure*}[!h]
\centering
\includegraphics[width=1.05\linewidth]{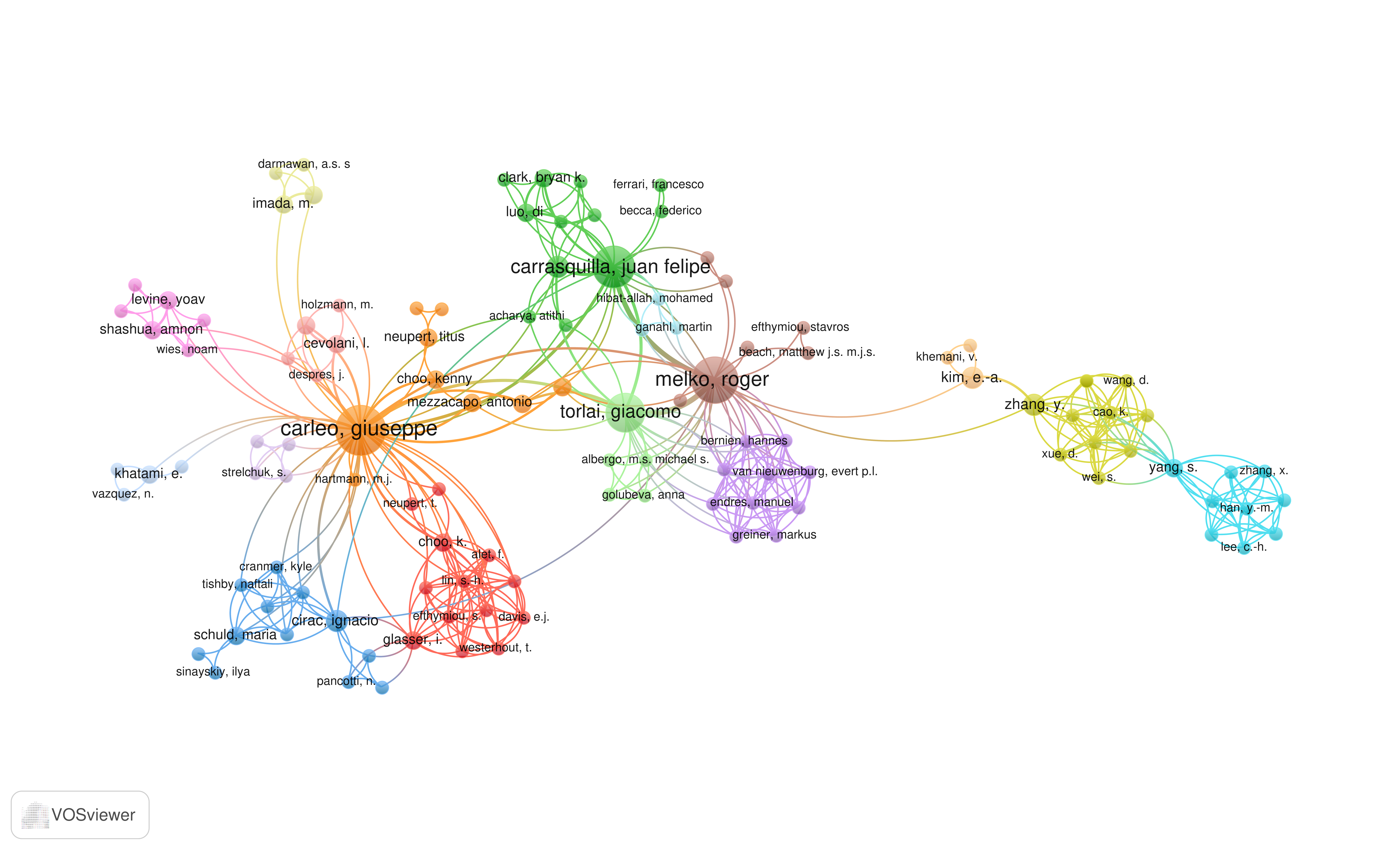}
\caption{Author map generated from the compiled bibliography. A cluster Between Carleo, Carrasquilla, Torlai, Cirac and Melko is appreciated, indicating collaboration between the Universities of Waterloo, Toronto, Complutense de Madrid and the Flatiron institute of New York. Mapping was limited to the biggest connected author cluster for visualization purposes.}
\label{fig:authors}
\end{figure*}

The search and compilation of the relevant literature was achieved via specialized search equations applied both in Scopus and Google Scholar. The used search equations are listed in table \ref{table:searchequations}. An author map (see figure \ref{fig:authors}) was generated from the compiled bibliography using VOSViewer. 
Initially, the generic search equations ("machine learning" AND "quantum physics") and ("Neural-Network Quantum States" OR  "Neural network quantum states" OR "Neural quantum states") were considered as a naive starting point. By parting from this search equation, taking \supercite{carleo_solving_2017} as a reference publication and iteratively refining the obtained results via keyword mapping, the remaining search equations (table \ref{table:searchequations}: 3-11) were subsequently derived. The study of the collected literature was achieved by means of an annotated literature scheme.

We remark the particular importance of Google Scholar in this work, given that this platform provides references both in indexed journals and in pre-print platforms such as the arXiv; with the latter being of key importance in the finding of recent literature on novel, rapidly-growing lines of research such as the one subject of this work.

\begin{table*}[!t]
\centering
\begin{tabular}{|l|p{16cm}|}
\hline
\# & \textbf{Search Equation}                                                                                                                                                                                                                         \\ \hline
1  & "Machine Learning" AND "Quantum Physics"                                                                                                                                                                                                         \\ \hline
2  & "Neural-Network Quantum States" OR  "Neural network quantum states" OR "Neural quantum states"                                                                                                                                                   \\ \hline
3  & (("Neural Networks" AND ("Autoregressive" OR "Recurrent")) AND ("wavefunctions" OR "wave functions" OR "quantum many body" OR "condensed matter OR "quantum phase transition" OR "quantum dynamics" OR "spin")                                   \\ \hline
4  & ("Machine Learning" OR "Deep Learning" "Neural Networks" OR "Restricted Boltzmann Machine") AND ("wavefunctions" OR "wave functions" OR "quantum many body" OR "condensed matter" OR "quantum phase transition" OR "quantum dynamics" OR "spin") \\ \hline
5  & ("Machine Learning" OR "Deep Learning" "Neural Networks" OR "Restricted Boltzmann Machine") AND (("spin" OR "fermion" OR "boson") AND ("chain" OR "lattice"))                                                                                    \\ \hline
6  & ("Machine Learning" OR "Deep Learning" "Neural Networks" OR "Restricted Boltzmann Machine") AND ("Ising" OR "Heisenberg" OR "Bose" OR "Hubbard")                                                                                                 \\ \hline
7  & ("Machine Learning" OR "Deep Learning" "Neural Networks" OR "Restricted Boltzmann Machine") AND (("Ising" OR "Heisenberg" OR "Bose" OR "Hubbard") AND (Localization OR Delocalization))                                                          \\ \hline
8  & ("Machine Learning" OR "Deep Learning" "Neural Networks" OR "Restricted Boltzmann Machine") AND ( ("quantum information" OR "quantum state tomography") OR ("spin" AND ("Chain" or "Lattice")) )                                                 \\ \hline
9  & ("Machine Learning" OR "Deep Learning" "Neural Networks" OR "Restricted Boltzmann Machine") AND ("spin" AND ("chain" OR "lattice") AND "dynamics")                                                                                               \\ \hline
10 & ("Machine Learning" OR "Deep Learning" "Neural Networks" OR "Restricted Boltzmann Machine") AND (("Ising" OR "Heisenberg" OR "Bose" OR "Hubbard") AND "dynamics")                                                                                \\ \hline
11 & ("Machine Learning" OR "Deep Learning" "Neural Networks" OR "Restricted Boltzmann Machine") AND ("Variational Montecarlo" OR "Variational Monte Carlo")   
\\ \hline
\end{tabular}
\caption{Identified search equations  for literature finding.}
\label{table:searchequations}
\end{table*}

\section{State of the art}



Since 2017, several lines of research focused on the study of the intersection between machine learning and quantum mechanics have surfaced in the form of several scientific publications that report significant advancements in computational physics. These contributions have enriched the current landscape of computational physics by taking advantage of both modern computational power and the rich theoretical and empirical advancements that have been developed during the past two decades and culminated in the so-called contemporary AI revolution.

Of the aforementioned contributions, the work by G. Carleo and M. Troyer~\supercite{carleo_solving_2017} can be regarded as one of the first modern publications on NQS. In this pioneering work, 
that coined the term `Neural-Network Quantum State', a paradigmatic generative model termed Restricted Boltzmann Machine (RBM) is used as a generic neural-network variational wavefunction ansatz (neural-network quantum state) in order to approximate the ground-state wavefunction of a quantum many-body system. The computational complexity of this RBM-based method scales quadratically with respect to the number of bodies, carrying this a significantly lower computational cost than a direct diagonalization scheme for the case of large many-body systems. Restricted Boltzmann Machines were initially conceived under the name Harmonium by P. Smolensky as part of his Harmony Theory of cognitive science\supercite{smolensky_information_1987}. They were later popularized by G. Hinton, who conceived an efficient algorithm for their training\supercite{hinton_training_2002}. RBMs consist of an input, visible unit layer and a latent, hidden unit layer, and previous to their application in quantum physics they were widely applied and studied for their dimensionality-reduction capabilities, and were usually trained by direct or indirect minimization of the Kullback-Leibler (KL) divergence between the probability distribution of the RBM and the one associated to training data. RBMs are physically equivalent to a system of two fully-connected layers of classical spins described by the canonical ensemble\supercite{torlai_learning_2016, torlai_giacomo_augmenting_2018}.

The NQS approximation to the ground state wavefunction is reached via an iterative optimization scheme of the energy functional based on a complex-valued generalization of the natural gradient method, known in the literature as Stochastic Reconfiguration (SR)\supercite{sorella_green_1998}. The SR method accelerates the convergence of variational Montecarlo methods compared to first order optimization algorithms such as stochastic gradient descent. Stochastic reconfiguration is, effectively, an imaginary time evolution over the variational subspace of the NQS wavefunction. Each step in the optimization process involves sampling from the RBM via Markov-Chain Metropolis Montecarlo (MCMMC), where the obtained set of samples effectively acts as a step-wise training set. For large enough systems, sampling cannot be performed in a exact way due to the intractable nature of the RBM partition function, the computation of which can be circumvented via the construction of the samples using the Metropolis algorithm. Recently, the linear method has also been considered as a feasible approach for optimizing NQS\supercite{Frank_Kastoryano_2021}.

\begin{figure}[!b]
\centering
\includegraphics[width=0.99\linewidth]{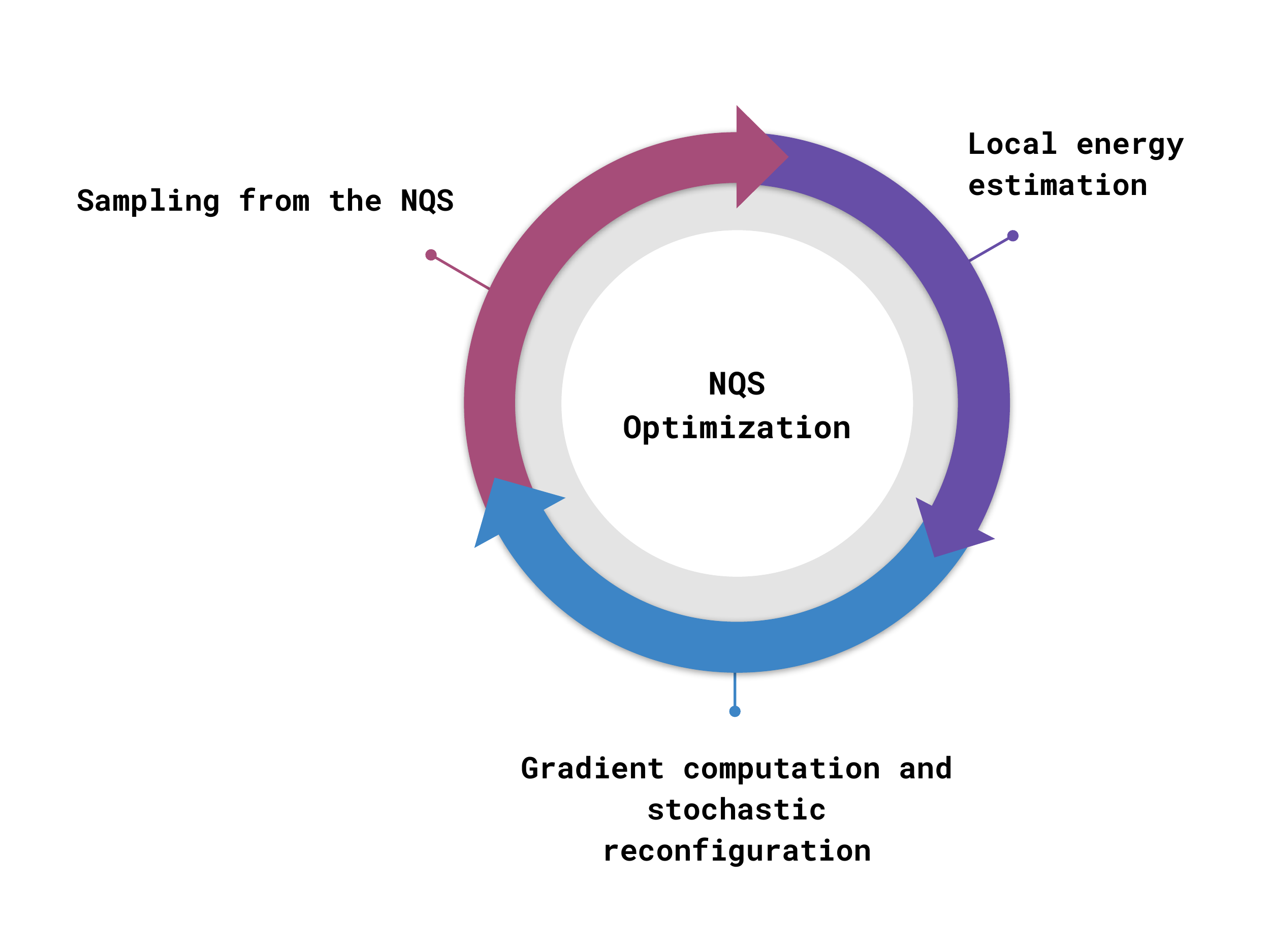}
\caption{General training scheme for the ground-state approximation of a NQS.}
\label{fig:nqs_training}
\end{figure}

\begin{figure}[!t]
\centering
\includegraphics[width=1.1\linewidth]{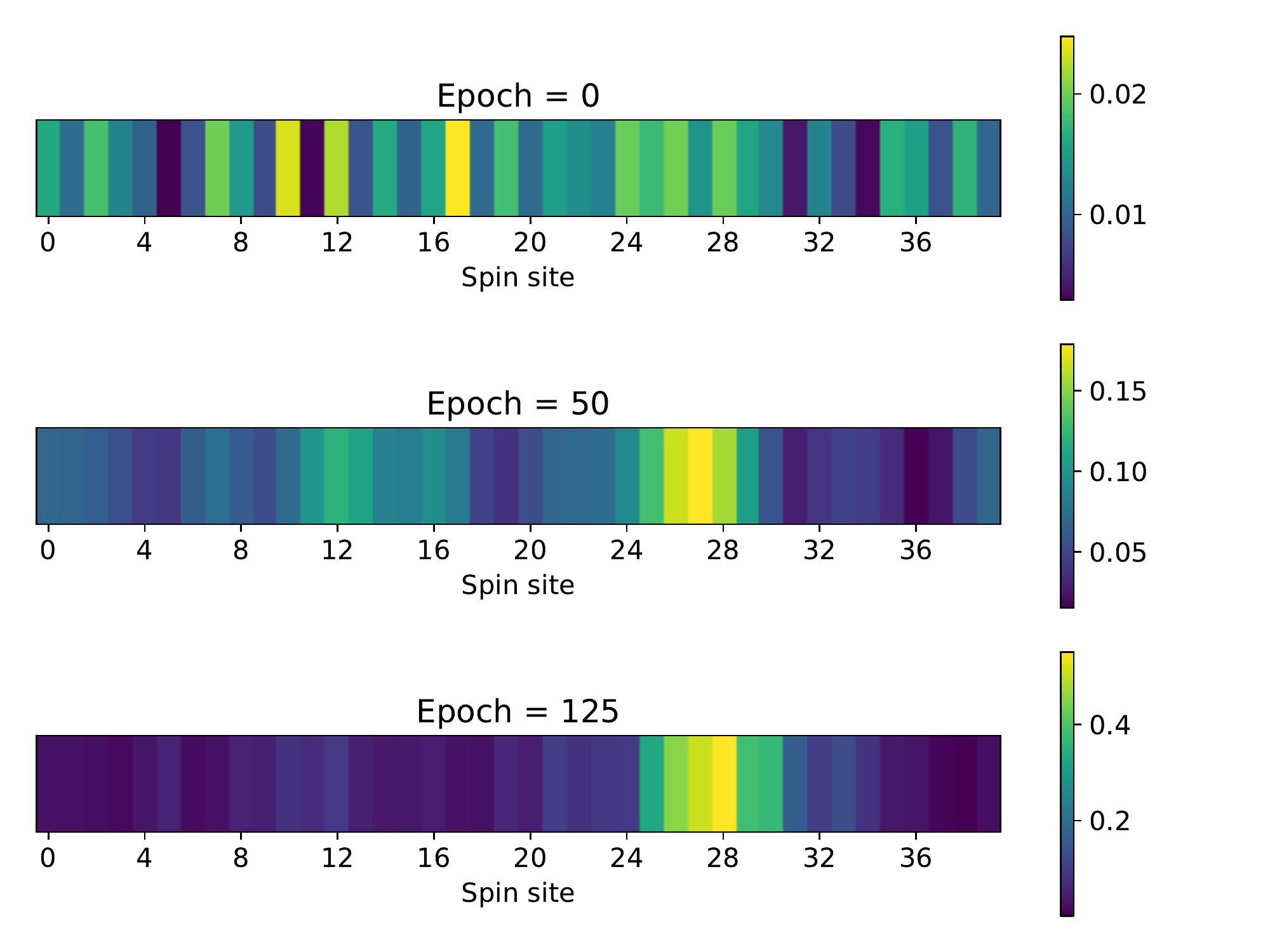}
\caption{Evolution of the norm of the weights connecting the visible layer with an arbitrary hidden neuron for a 1D transverse field Ising model ground-state wavefunction described by an RBM NQS. At the end of the training, weights connecting the chain with individual hidden neurons effectively act as local filters over different regions of the chain, reflecting the short-range interactions of this particular Hamiltonian. Given that fully-connected NNs such as RBMs are not aware of the relative position of its inputs (in comparison to CNNs), this result shows that the model naturally learns the relative position of the spin configurations during training. The  results are  reproduced following~\supercite{carleo_solving_2017}.}
\label{fig:weights}
\end{figure}

Following the sampling process, an stochastic estimate of the energy functional gradients are computed from the samples. The final step of each training epoch involves the solution of a linear equation system, as required by SR. The dimensionality of such system grows linearly with the number of variational parameters of the RBM, and its solution can be obtained via direct inversion (or pseudoinversion) of the SR Matrix or via iterative Krylov subspace methods such as conjugate gradient, with the latter being already successfully connected with SR in \supercite{neuscamman_optimizing_2012}. The process described above is repeated until reaching a specified convergence threshold. The work by Carleo and Troyer also demonstrates the capability of executing unitary time evolution over the approximated ground-state, thus supporting time-dependent variational Montecarlo. Although SR offers faster convergence, Adam optimizer \supercite{kingma_adam_2017} has also proven capable of optimizing NQS \supercite{carleo_netket_2019, sharir_deep_2020}. The described scheme is illustrated in figure \ref{fig:nqs_training}. In reference \supercite{carleo_solving_2017} it is shown that, after training, each neuron of the RBM acts as an effective filter over a region of the spin chain or lattice,  in a similar fashion to convolutional neural networks; a reproduction of this result is illustrated in figure \ref{fig:weights}. 

Another high-impact result, published in parallel to the aforementioned, introduces a Quantum State Tomography (QST) protocol base on neural-network quantum states\supercite{torlai_neural-network_2018}. This novel approach, regarded as Neural-Network Quantum State Tomography (NQST), uses a neural-network to reconstruct the wavefunction of a quantum system given a set of configuration measurements or samples. The involved computational methodology minimizes the KL divergence between the probability distribution associated to the samples and the probability distribution defined by the squared norm of the NQS wavefunction, with the added difficulty that such wavefunction is a complex-valued probability amplitude. The reported optimization scheme uses a complex-valued RBM trained via ordinary stochastic gradient descent for its radial part and natural gradient descent for its complex-phase. The reconstructed wavefunction was also proven capable of executing unitary dynamics. Following this work, subsequent contributions have improved the precision of the method \supercite{torlai_precise_2019, palmieri_experimental_2020}; extended it to the density-matrix formalism\supercite{carrasquilla_reconstructing_2019}; used recurrent neural networks as an alternative NQST architecture\supercite{torlai_integrating_2019,torlai_precise_2020, Morawetz_De_Vlugt_Carrasquilla_Melko_2021}; extended the protocol to local-measurement-based QST\supercite{xin_local-measurement-based_2019}; and even proved the capabilities of powerful generative models such as variational auto-encoders\supercite{kingma_auto-encoding_2014} to reconstruct complex many-body wavefunctions\supercite{rocchetto_learning_2018,luchnikov_variational_2019,walker_deep_2020}. 


Despite sampling methods based on the Metropolis algorithm allow the construction of samples from otherwise intractable probability distributions, these methods are prone to exhibit both high correlation between samples, and also the inability to provide adequate samples when approaching energy barriers, such as the ones usually present in certain regimes of quantum matter. 

\begin{figure}[!t]
\centering
\includegraphics[width=0.90\linewidth]{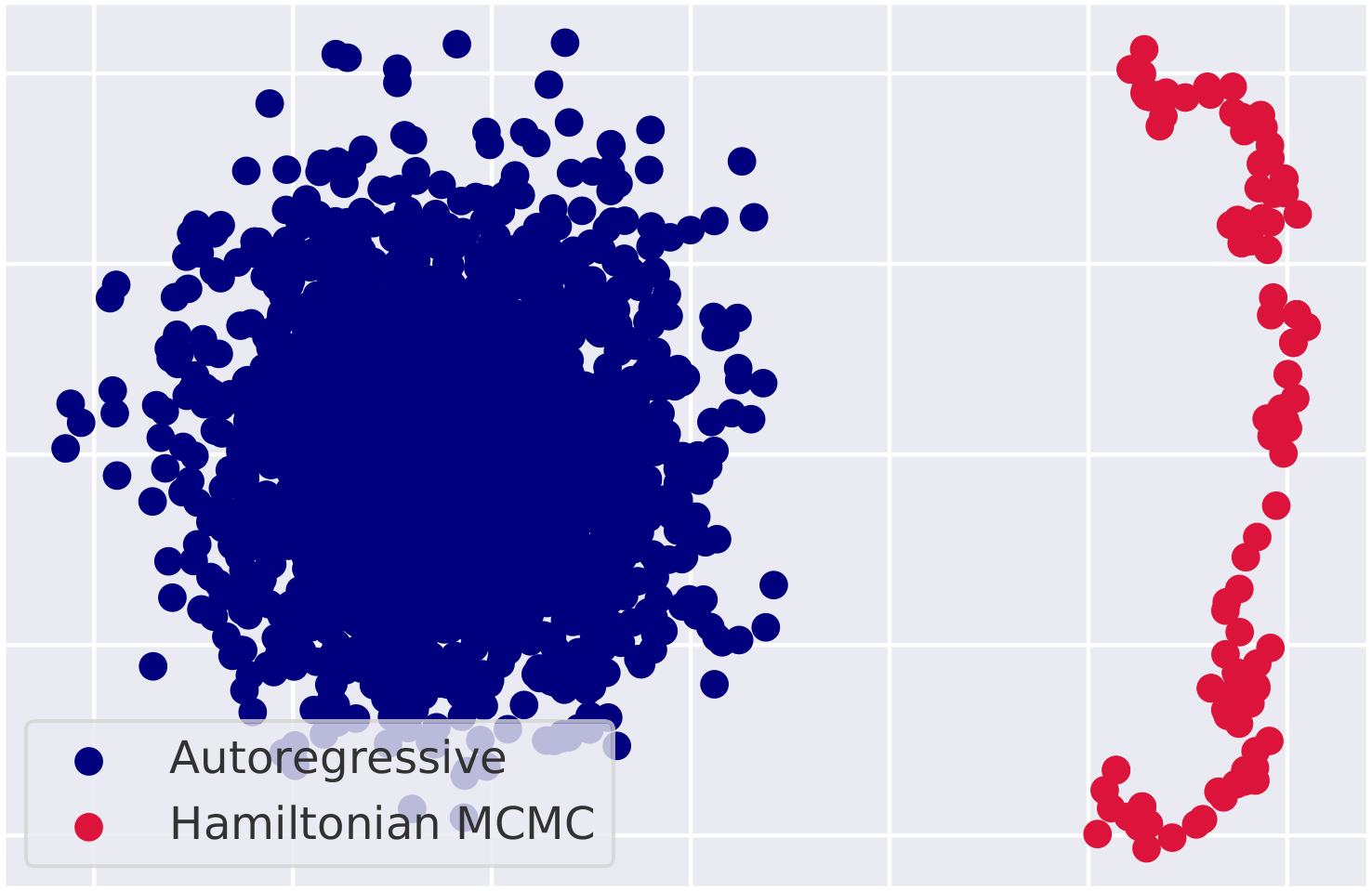}
\caption{Principal component analysis of samples obtained both by autoregressive sampling and a Hamiltonian MCMC chain for the ground-state of a 100-long transverse-field Ising chain in the disordered phase, represented by an autoregressive transformer. The uncorrelated samples of the autoregressive sampler fall inside what can be regarded as a representative region of the ground state of the system, while the Hamiltonian MCMC samples are confined to a less-diverse chain on a distant starting region of the configuration space.}
\label{fig:PCA}
\end{figure}

Neural Autoregressive Density Estimators (NADEs)\supercite{uria_neural_2016}, popularized by Google Deepmind\supercite{oord_conditional_2016, oord_wavenet_2016}, are generative neural-networks that define a probability distribution from which sampling can be performed exactly. Given that the connections between layers of a NADE are causal, sampling can be performed via the chain rule of conditional probabilities, thus eliminating the need of using construction methods such as MCMMC. A recent contribution~\supercite{sharir_deep_2020} has extended the probabilistic framework that supports NADEs to the notion of a complex-valued probability amplitude, from which a physical ground-state wavefunction can be obtained. This approach completely removes the need of using MCMMC for NQS representation and provides uncorrelated samples with higher computational efficiency: in figure \ref{fig:PCA} we illustrate via principal component analysis how uncorrelated autoregressive samples naturally fall into a representative region of the configuration space, while the MCMMC samples are rather spread along a correlated path. Another recent contribution by Hibat {\it et al.}\supercite{hibat-allah_recurrent_2020} reports the use of a recurrent NQS architecture, which also supports exact sampling via the autoregressive property, to solve prototypical systems of up to 1000 spins with a remarkably low amount of variational parameters and with the capability of estimating Renyi Entropies\supercite{wang_calculating_2020}. 

Further works related to NQS ground-state research include a construction method for the classical representation of quantum states via Deep Boltzmann Machines\supercite{carleo_constructing_2018}, an error\supercite{yang_approximating_2019}, learnability scaling\supercite{sehayek_learnability_2019} and representability\supercite{yang_neural_2020} analysis of NQS under different tensor operations, an expressive power analysis on different spins, Hamiltonians and architectures \supercite{cai_approximating_2018}, a study of the expressive power of deep NQS \supercite{Sharir_Shashua_Carleo_2021} compared to variational tensor networks, an application to the ground-state simulation of few-body bosonic systems\supercite{saito_method_2018}, an analysis of the limitations of shallow NQS architectures\supercite{gao_efficient_2017} and RBMS\supercite{lu_efficient_2019} as universal ground-state approximators, a detailed analysis of the transverse-field Ising model\supercite{shi_neural-network_2019}, a random-sampling approach to the solution of the eigenvalue problem\supercite{liu_random_2020}, a pair-product approach to compliment the expressive power of RBM-based NQS\supercite{nomura_restricted_2017}, an study of the connection between RBMs and tensor-network from a representability perspective\supercite{glasser_neural-network_2018}, a deep study of the entanglement representation capabilities of convolutional neural-networks\supercite{levine_deep_2018}, the ground state approximation\supercite{saito_solving_2017, choo_symmetries_2018}, phase boundary exploration\supercite{mcbrian_ground_2019} of paradigmatic bosonic systems such as the Bose Hubbard Model, the use of group-equivariant convolutional neural-networks to optimize NQS ground-state approximation\supercite{Roth_MacDonald_2021}, the construction of gauge-invariant autoregressive NQS\supercite{Luo_Chen_Hu_Zhao_Hur_Clark_2021}, the parametrization of electronic wavefunctions by means of autoregressive NQS\supercite{Barrett_Malyshev_Lvovsky_2021}, a super-resolution approach to extrapolate spin configurations of the transverse-field Ising model to bigger lattices\supercite{efthymiou_super-resolving_2019}, the iterative retraining of Recurrent NQS to progressively scale to larger systems\supercite{Roth_2020}, the study of disordered quantum Ising chains via dense and sparse RBMs\supercite{Pilati_Pieri_2020}, the approximation of the ground-state of the frustrated $J1-J2$ Heisenberg model by means of a dual NQS approach\supercite{Bukov_Schmitt_Dupont_2021}, the smmulation of spin liquids on a Honey lattice via RBMs\supercite{Li_Yang_Xu_2021} and, lastly, the use of graph neural-networks to learn a parametrized structured variational manifold NQS that achieves state-of-the-art results on quantum many-body benchmarks\supercite{Kochkov_Pfaff_Sanchez-Gonzalez_Battaglia_Clark_2021}. 

The NQS unitary dynamics algorithm introduced in \supercite{carleo_solving_2017} is, computationally speaking, a problem similar to SR, but demands a higher number of samples and limits the possibility of regularizing the involved linear equation system, implying this a higher computational cost and the surfacing of new challenging numerical instabilities. The dynamics simulation capabilities of NQS have been explored and extended in several subsequent works. In \supercite{schmitt_quantum_2019}, dynamics over up to $8\times 8$ transverse-field Ising lattices are explored and compared to iPEPS\supercite{jahromi_universal_2019} results, with the authors reporting the numerous computational techniques and workarounds used in order to ensure the stability and accuracy of the time evolution. Another contribution \supercite{lopez-gutierrez_real_2019} proposes the use of an alternative algorithm based on the solution of the time-dependent Schrodinger equation with an NQS wavefunction via the midpoint-method; this novel approach involves the training of a new NQS at each time step $t+1$ by minimizing a cost function with respect to the $t$ time step wavefunction over a set of samples, thus allowing the possibility of changing and adapting the NQS architecture on the fly, with this being specially useful in cases where certain physical regimes reached during the time evolution require a higher expressive power NQS. In \supercite{zhang_predicting_2020} it is shown that a simple recurrent neural-network is able to predict the time evolution of a one-dimensional transverse-field Ising model, and also scale to larger systems after training. In \supercite{carleo_unitary_2017} NQS unitary dynamics simulation over strongly-interacting Bose gases in continuous space is executed. Reference\supercite{lee_neural-network_2020} introduces a novel quantum algorithm based on the use of a modified RBM for the variational simulation of closed and open many-body dynamics, while circumventing the vanishing gradient problem. In \supercite{hendry_machine_2019}, a method for the many-body dynamical simulation on the frequency domain is devised. NQS have also proven capable of simulating the dynamics of open quantum systems, with \supercite{hartmann_neural-network_2019} introducing a first approach for small systems and \supercite{luo_autoregressive_2020} further improving the previous results with the use of autoregressive NQS. In \supercite{Burau_Heyl_2021}, a quantum renormalization group approach is combined with neural-networks to compute the long-time coherent dynamics of large many-body localized systems. In \supercite{yoshioka_constructing_2019}, stationary states of open quantum many-body systems have been approximated using NQS and several physical observables have been computed. Finally, in \supercite{Reh_Schmitt_Gärttner_2021} a simulation of up to 40 spins in a dissipative Heisenberg lattice in one and two dimensions has been reported.

NQS have also been successfully applied to the analysis, exploration and discovery of physical phases; including the finding of quantum critical points\supercite{zen_finding_2020-1}, phase-diagram reconstruction of the Bose-Hubbard model\supercite{vargas-calderon_phase_2020}, approximation of states on chiral topological phases\supercite{kaubruegger_chiral_2018, deng_machine_2017},  the study of entanglement transitions by means of a RBM\supercite{Medina_Vasseur_Serbyn_2021} and representation capability analysis near quantum criticality\supercite{morningstar_deep_2018, czischek_quenches_2018}. Although state configurations (i.e. spin configurations) can be regarded as relatively simple computational data, they are a reflection of highly-complex phenomena that takes place inside the Hilbert space of a quantum many-body system. Being fed by such configurations, fedforward neural-networks have also proven capable of identifying phases of matter, phases transitions and order parameters from raw-spin configurations on a wide variety of Hamiltonians following a supervised learning approach\supercite{carrasquilla_machine_2017, venderley_machine_2018, bohrdt_classifying_2019, richter-laskowska_machine_2018, rem_identifying_2019, ohtsuki_drawing_2020, liu_discriminative_2018, broecker_machine_2017, bachtis_mapping_2020, ohtsuki_deep_2017, schindler_probing_2017, berezutskii_probing_2020, theveniaut_neural_2019, schulz-beach_explorations_2020, zejian_machine_2019,pancotti_methods_nodate, Huang_Kueng_Torlai_Albert_Preskill_2021}. Alternative machine learning algorithms have also proven highly capable of handling physical phases of matter\supercite{van_nieuwenburg_learning_2017, zhang_interpretable_2019, vargas-hernandez_extrapolating_2018, shirinyan_self-organizing_2019, ponte_kernel_2017, hu_discovering_2017, holanda_machine_2020, dong_machine_2019, che_topological_2020, canabarro_unveiling_2019, yang_visual_2020, schulz-beach_explorations_2020}.


\section{Final Remarks and Perspectives}

Although recent contributions have reported the optimization of NQS up to $10^3$ bodies\supercite{hibat-allah_recurrent_2020}, successful NQS dynamics are still in the order of $10^1$ bodies. A main perspective is the development of novel methodologies and/or computational techniques that make possible the study of the dynamics of larger systems. Current advances in NN architectures such as self-attention mechanisms\supercite{Vaswani_Shazeer_Parmar_Uszkoreit_Jones_Gomez_Kaiser_Polosukhin_2017}, Hamiltonian NNs and graph NNs open interesting perspectives for the further development of NQS architecture research.

Neural-networks have also proven capable of approximating ground-states of arbitrary sparse Hamiltonians in continuous spaces\supercite{teng_machine-learning_2018}. Current state of this approach provides the advantages of a universal ansatz but is, however, still penalized by the quantum curse of dimensionality. A more efficient approach is yet to be devised. 

With the current development of hardware specifically designed for deep learning purposes, a new perspective for the representation of NQS on such specialized hardware arises. This proposal was already analyzed in \supercite{czischek_sampling_2019, czischek_neural-network_2020}, which identifies both advantages and difficulties. A practical implementation remains to be executed.

Despite not being capable of providing explicit probability distributions as needed for NQS ground-state approximation and NQST, Generative Adversarial Networks (GANs) have found applications in many-body physics, by introducing an approach for modelling physical phases of quantum spin systems with these kind of models\supercite{singh_generative_2020}. A wider spectrum of applications for these powerful generative models remains to be seen.


As illustrated in this work, the diverse applications of Neural-network Quantum States in many-body quantum physics have configured a rapidly-growing, high-impact field of research that has vastly expanded the current landscape of computational physics, condensed matter, and quantum information science. The recent birth of this formalism together with its rapid advancements opens promising perspectives both for new computational techniques to come and the maturity of the already present ones.

\section{Acknowledgments}

This work was supported by  Vicerrectoría de Investigaciones at Universidad del Valle (Grant CI 71212).

\printbibliography


\end{document}